\documentclass[aps,superscriptaddress,superbib,twocolumn,groupedaddress,letterpaper]{revtex4}

\usepackage{natbib}
\bibpunct{[}{]}{,}{n}{}{}

\usepackage{amsmath,amssymb}
\usepackage{graphicx}

\usepackage{units}
\usepackage{journames}

\usepackage[final]{neel}

\usepackage{hyperref}
\definecolor{link_green}{rgb}{0.0,0.7,0.0}
\hypersetup{pagebackref,breaklinks,citecolor=link_green,colorlinks,bookmarks,bookmarksopen,
  pdfauthor="Olivier Fruchart",pdfstartview=FitH}
\def\figurename{Fig.}
\usepackage[all]{hypcap}

\begin{document}

\renewcommand\topfraction{0.8}
\renewcommand\bottomfraction{0.7}
\renewcommand\floatpagefraction{0.7}

\def\Ht{H_{\mathrm t}}%

\title{Controlled switching of N\'{e}el caps in Bloch magnetic domain walls}%

\author{F.~Cheynis}
\affiliation{Institut N\'EEL, CNRS \& Universit\'{e}
Joseph Fourier -- BP166 -- F-38042 Grenoble Cedex 9 -- France}%
\affiliation{Institut National Polytechnique de Grenoble -- France}%
\author{A.~Masseb\oe uf}
\affiliation{CEA-Grenoble, INAC/SP2M/LEMMA, 17 rue des Martyrs, Grenoble, France}%
\author{O.~Fruchart}
\email[]{Olivier.Fruchart@grenoble.cnrs.fr}\affiliation{Institut N\'EEL, CNRS \& Universit\'{e}
Joseph Fourier -- BP166 -- F-38042 Grenoble Cedex 9 -- France}%
\author{N.~Rougemaille}%
\affiliation{Institut N\'EEL, CNRS \& Universit\'{e}
Joseph Fourier -- BP166 -- F-38042 Grenoble Cedex 9 -- France}%
\author{J.~C.~Toussaint}
\affiliation{Institut N\'EEL, CNRS \& Universit\'{e}
Joseph Fourier -- BP166 -- F-38042 Grenoble Cedex 9 -- France}%
\affiliation{Institut National Polytechnique de Grenoble -- France}%
\author{R.~Belkhou}
\affiliation{Synchrotron SOLEIL, L'Orme des Merisiers Saint-Aubin - BP 48, F-91192 Gif-sur-Yvette Cedex, France}%
\affiliation{ELETTRA, Sincrotrone Trieste, I-34012 Basovizza, Trieste, Italy}%
\author{P.~Bayle-Guillemaud}
\affiliation{CEA-Grenoble, INAC/SP2M/LEMMA, 17 rue des Martyrs, Grenoble, France}%
\author{A.~Marty}
\affiliation{CEA-Grenoble, INAC/SP2M/NM, 17 rue des Martyrs, Grenoble, France}%

\date{\today}


\begin{abstract}

While magnetic hysteresis usually considers magnetic domains, the switching of the core of
magnetic vortices has recently become an active topic. We considered Bloch domain walls, which are
known to display at the surface of thin films flux-closure features called N\'{e}el caps. We
demonstrated the controlled switching of these caps under magnetic field, occurring via the
propagation of a surface vortex. For this we considered flux-closure states in elongated
micron-sized dots, so that only the central domain wall can be addressed, while domains remain
unaffected.

\end{abstract}

\maketitle

\vskip 0.5in

\vskip 0.5in


Data storage relies on the handling of two states, called bits. In magnetoelectronic devices bits
are stored using the two directions of magnetization in nanoscale bistable domains. In hard-disk
drives these are written in granular continuous thin films, while in solid-state Magnetic Random
Access Memories (MRAMs) bits rely on dots patterned with lithography\cite{bib-CHA2007}. While
miniaturisation is the conventional way to fuel the continuous increase of device density,
disruptive solutions are also sought. To these belong in recent years many fundamental studies
investigating no more the domains, but the manipulation of domain walls and magnetic vortices as
objects in themselves. The degree of freedom associated with these may be their location, \eg for
a domain wall moved along a stripe\cite{bib-ALL2005}, or their internal structure, \eg the
up-or-down polarity of a vortex in a circular dot\cite{bib-OKU2002,bib-VAN2006,bib-YAM2007}. In
the latter case the chirality of the flux-closure around the vortex might also be exploited as a
second degree of freedom\cite{bib-ZHU2000,bib-KIK2001,bib-ARR2002}. In this context it is of
fundamental interest to revisit all existing internal degrees of freedom in classical domain
walls, and determine whether or not they might be switched independently of their environnement.

\begin{figure}
  \begin{center}
  \includegraphics[width=83mm]{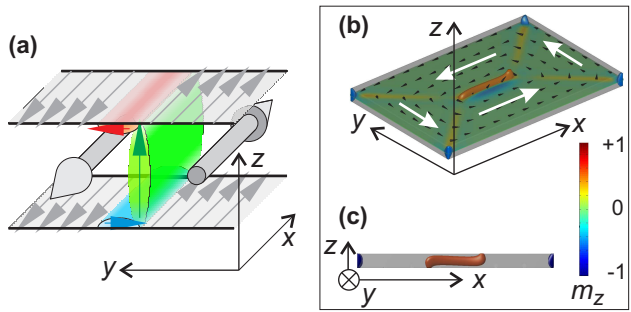}%
  \caption{\label{fig-walls}(a)~Schematic view of a domain wall of Bloch type, terminated by a N\'{e}el cap at each surface.
   The magnetization is opposite in the top and bottom N\'{e}el caps, and points along the $\pm y$ axes
    \ie across the plane of the wall. (b-c)~3D and cross section views of a so-called \textsl{Landau} state in a rectangular magnetic dot ($\lengthnm{700\times500\times50}$). In (b)~only volumes with normalized perpendicular magnetization $|m_z|>0.5$ are displayed, which highlights the vortex and the domain wall. The map of $m_z$ at mid-height is also shown, and the in-plane curling of magnetization is indicated by white
  arrows.}
  \end{center}
\end{figure}

\begin{figure}
  \begin{center}
  \includegraphics[width=83mm]{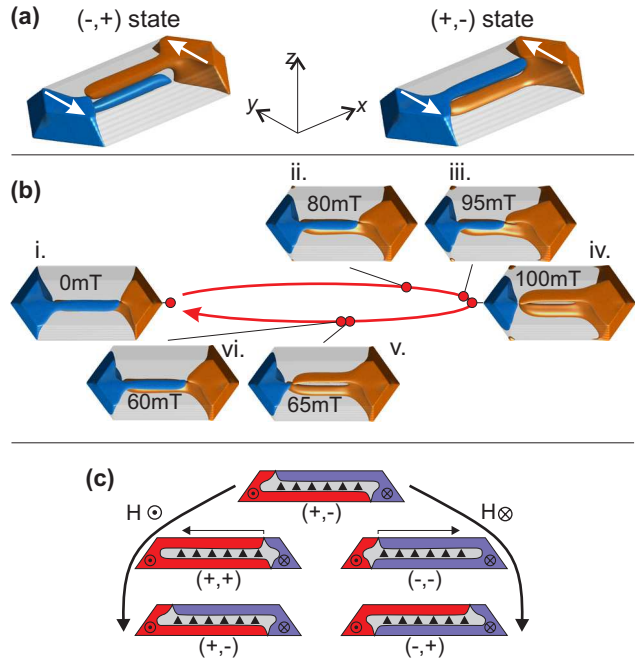}%
  \caption{\label{fig-hysteresis}(a)~Simulation in an elongated dot of the $(-,+)$ and $(+,-)$ states. Only volumes with $|m_y|>0.5$ are displayed, while positive and negative values appear red and blue, respectively.
   (b)~A magnetization process of N\'{e}el caps under $H_\mathrm{t}>0$
   (c)~Schematic cross-sectional view of the switching of NCs:
   the final state is $(-,+)$ or $(+,-)$
   depending on the sign of the applied field.}
  \end{center}
\end{figure}

In this Letter we consider an already known internal degree of freedom in Bloch domain walls, the so-called N\'{e}el caps~(NCs), and report the controlled magnetic switching of these NCs. Bloch walls in thin films have been extensively addressed\cite{bib-HUB1969,bib-LAB1969}. In contrast to the simplest text-book case where a translational invariance is assumed within the DW plane, Bloch walls in thin films
with in-plane magnetization are known to display an extra degree of freedom beyond the up-or-down
polarity of its core: the direction of magnetization in the NCs, occurring at both surfaces of the
wall to decrease the magnetostatic energy\bracketsubfigref{fig-walls}a. The name NC was given only
later\cite{bib-FOS1996}, as the arrangement of magnetization in NCs mimics that of a N\'{e}el wall, a type of DW with solely in-plane magnetization\cite{bib-NEE1955}. At remanence the two NCs are antiparallel to each other and
thus in principle two degenerate ground states exist, which we named after the direction of
magnetization of the bottom and top NCs, \ie $(-,+)$ and
$(+,-)$\bracketsubfigref{fig-hysteresis}a. This overall arrangement was named an asymmetric Bloch
wall\cite{bib-LAB1969}. So far nobody demonstrated that the remanent state of
antiparallel NCs could be selected reliably.
This mainly stems from the fact that in thin films, where such processes had been sought,
the magnetization in the domains rotates easily under applied fields, often affecting the location and
even the existence of domain walls\cite{bib-HUB1998b}. To lift this limitation we considered elongated dots
displaying a flux-closure state, were a finite-size Bloch domain wall is
stabilized by the internal dipolar fields\bracketsubfigref{fig-walls}{b-c}. The structure of these
flux-closure states has long since been described, and is now called a Landau
state\cite{bib-LAN1935,bib-ARR1979b,bib-ARR1997,bib-HER1999,bib-FRU2003c,bib-FRU2005}. Here we
demonstrate numerically and experimentally that NCs can be switched controllably under the
application of a magnetic field without affecting the other two degrees of freedom found in the
dot: the core with perpendicular magnetization of the Bloch wall, nor the chirality of the
in-plane flux-closure.

The sample consists of epitaxial micron-sized self-assembled elongated Fe$(110$) dots with
atomically-smooth facets. These were fabricated using UHV pulsed-laser deposition on
$\saphir(11\overline20)$ wafers buffered with a layer of W$(110)$. The dots grow
spontaneously as a result of a dewetting process named Stranski-Krastanov growth mode,
driven by the large lattice mismatch between Fe and W~($\unit[\approx10]{\%}$). The facets arise
spontaneously and reproducibly with well-defined directions in relation with the
single-cristallinity of the dot\cite{bib-FRU2007}. The growth
temperature~(\tempC{530}) and nominal thickness~(\lengthnm{5}) were chosen so that the average
length, width and height of the dots are $\unit[1]{\micron}$, $\lengthmicron{0.5}$ and
$\lengthmicron{0.1}$, respectively. The dots were capped \insitu with $\lengthnm{0.7}$-thick Mo
followed by a $\lengthnm{2.5}$-thick Au layer to prevent oxidation during the transfer in air
between the growth chamber and the magnetic imaging setups\dataref{Echantillon FC28}.

Two high-resolution imaging instruments were used, bringing complementary informations. The first
instrument is an Elmitec GmbH LEEM/PEEM microscope (LEEM~V), based at the nanospectroscopy
beamline of ELETTRA synchrotron (Trieste,Italy). In the Low Energy Electron Microscopy (LEEM)
mode\cite{bib-BAU1994}, topographical features are revealed with a lateral resolution below
$\lengthnm{10}$. In the X-ray Photo-Emission Electron Microscope mode~(XPEEM) the X-rays energy
was tuned at the Fe $\mathrm{L}_3$ edge, $h\nu=\unit[707]{eV}$. The magnetic contrast is obtained
using XMCD-PEEM (X-Ray Magnetic Circular Dichroism)\cite{bib-STO1999}. The spatial resolution is
here $\lengthnm{25}$, with a probing depth of $\lengthnm{10}$. Maps of the in-plane component of
surface magnetization parallel to the X-ray beam are thus achieved. As a second instrument we used
a FEI Titan transmission electron microscope in the Fresnel contrast mode\cite{bib-CHA1999}. The
microscope is equipped with a Gatan Imaging Filter for zero-loss filtering and a dedicated Lorentz
lens, and was operated at 300kV. We tilted the sample by $\angledeg{20}$ so that the field
produced by the Lorentz exhibits an in-plane component, which we used in our experiments to switch the
NCs. The Fresnel images highlight the location of vortices and domain walls separating the
in-plane domains, with a bright or dark contrast contrast revealing the
chirality of the domains around them, not the magnetization of their core.
The Sapphire substrate was mechanically polished and then ion-milled to permit transmission of the
electrons.

The micromagnetic simulations were performed using GL\_FFT, a custom-developed
code\cite{bib-TOU2002,bib-FRU2004c} based on a finite-differences scheme. The prism cell dimensions are
$\Delta_x=\Delta_y=\unit[3.9]{nm}$ and $\Delta_z=\unit[3.1]{nm}$. Due to the large thickness of
the dots bulk magnetic properties of Fe at $\tempK{300}$ have been used: magneto-cristalline
fourth-order anisotropy constant $K_1=\unit[\scientific{4.8}{4}]{J/m^{3}}$, exchange energy
$A=\unit[\scientific{2}{-11}]{J/m}$ and spontaneous magnetization
$\Ms=\unit[\scientific{1.73}{6}]{A/m}$.

\begin{figure}
  \begin{center}
  \includegraphics[width=82mm]{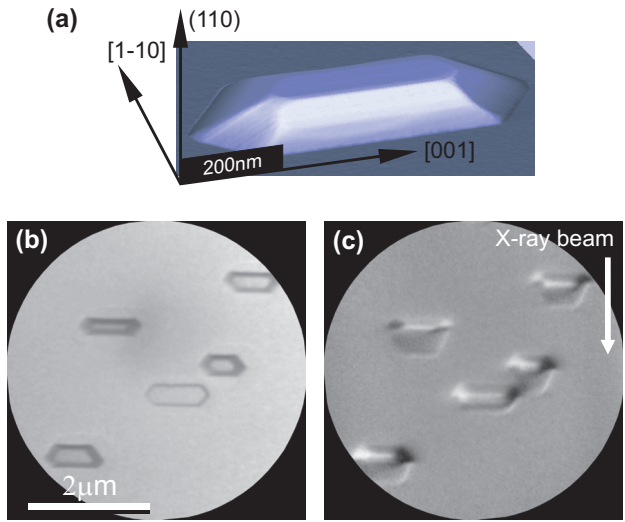}%
  \caption{\label{fig-up-to-down}(a)~3D view of a typical self-assembled epitaxial Fe(110)
  dot (Atomic force microscopy, true aspect ratios)
  \dataref{1129S005.hdf, Fabien}.
  (b)~LEEM (topography) and (c)~XMCD-PEEM (magnetism) typical view of an ensemble of dots.
  After magnetization at $\unit[-130]{mT}$ the dots are in the $(-,+)$ state at
   remanence whatever their size, height or aspect ratio. The white arrow indicates the direction of the X-rays, thus the component
   of in-plane surface magnetization imaged.}
  \end{center}
\end{figure}

We first report simulations. First notice on \figref{fig-walls}{b-c}, as already
known\cite{bib-ARR1979b,bib-SCH1991b}, that a vortex and a Bloch wall in a dot are topology equivalent. Conceptually the asymmetric Bloch wall can be derived from a vortex by shearing apart its top and bottom extremities. This leads to the prototypical Landau
structure\cite{bib-LAN1935}. In the following we call these extremities \textsl{surface
vortices}. We now focus on faceted dots\bracketfigref{fig-hysteresis}, whose shape and size are
those investigated experimentally. For a given chirality and polarity of the DW the set of NCs gives rise to two degenerate ground states at zero external field, \ie $(-,+)$ and
$(+,-)$\bracketsubfigref{fig-hysteresis}{a}. Starting from a $(+,-)$ state, a
transverse magnetic field $\Ht$ is applied (\ie along~$y$). Upon applying a positive $\Ht$ no
significant change occurs up to \unit[95]{mT}\bracketsubfigref{fig-hysteresis}{b}. At
$\Ht=\unit[100]{mT}$ the top surface vortex propagates through the top NC to the other end of the DW,
and settles atop the bottom vortex surface. The two NCs are then parallel and aligned along the
field direction~[$(+,+)$ state]. This arrangement is known as an \textsl{asymmetric N\'{e}el wall},
consistent with the finding in thin films that asymmetric N\'{e}el walls are favored against
asymmetric Bloch walls under a transverse field\cite{bib-MID1963,bib-HUB1998b}. Upon decreasing
$\Ht$ back to zero no significant change occurs down to $\unit[60]{mT}$, where suddenly the
surface vortex travels back to its initial position~[$(+,-)$ state]. Under now decreasing negative
fields the top NC remains unaffected, while it is the bottom NC that switches still around
$\Ht=\unit[-100]{mT}$\bracketsubfigref{fig-hysteresis}c [$(-,-)$ state]. Interestingly, when $\Ht$
is reduced back to zero it is again the top NC that switches back. This leaves a $(-,+)$ state at
remanence, \ie opposite to the initial state. The fact that it is always the top NC that switches
back when the field magnitude is reduced might be related to the tilted
facets\bracketsubfigref{fig-up-to-down}{a-b}, which break the symmetry between the top and bottom
surfaces. Thus owing to this breaking of symmetry and according to micromagnetic simulations, one
should be able to switch the set of NCs by applying a magnetic field transverse to the dot. At
remanence the top NC should be antiparallel to the applied
field\bracketsubfigref{fig-hysteresis}{c}.

The micromagnetic predictions were confirmed experimentally. As no significant magnetic field can
be applied while imaging with XMCD-PEEM, the state of the NCs was checked at remanence after
\exsitu application of a transverse magnetic field $\Ht$ with a given sign and magnitude. For each
field several tens of dots were imaged~(20 to 40). NCs are revealed as a thin stripe of dark or
light contrast along the length of the dots\bracketsubfigref{fig-up-to-down}c. In principle this
contrast could be mistaken as arising from a N\'{e}el wall. This is however ruled out as the magnetic
force microscopy signature of these walls is monopolar, while N\'{e}el walls would induce bipolar
contrasts\cite{bib-FRU2003c}. We now describe the results. In the as-grown state the $(-,+)$ and
$(+,-)$ states were found in equal ratios within statistical fluctuations. In contrast the
occurrence of the $(+,-)$ state [$(-,+)$, resp] reaches \unit[95]{\%} after application of
$\Ht=\unit[+150]{mT}$ ($\Ht=\unit[-150]{mT}$, resp.). The mean switching field is
$\Hsw=\unit[120]{mT}$ with a $\unit[\pm10]{mT}$ distribution. This value is in close agreement
with the numerical simulations presented above~($\Hsw=\unit[100]{mT}$). The remaining
$\unit[5]{\%}$ of dots are still found in the $(-,+)$ [resp. $(+,-)$] state. This lack of
switching could result from subtle changes in the arrow-shape of the dots such as an asymmetry,
which may induce the departure of the bottom surface vortex at decreasing field, instead of the
top one. This remains to be studied.

\begin{figure}
  \begin{center}
  \includegraphics[width=80mm]{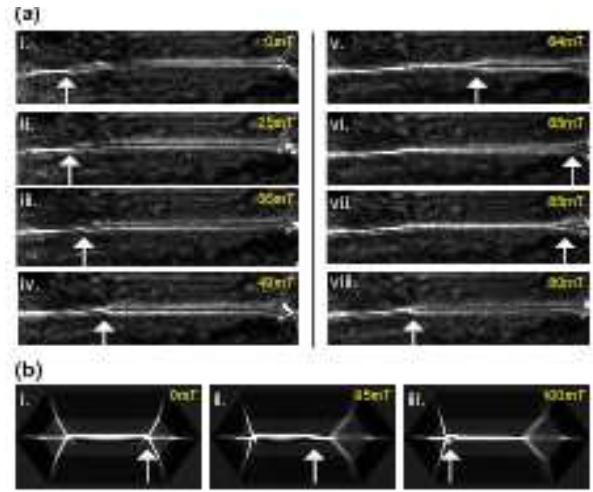}%
  \caption{\label{fig-lorentz}(a)~Fresnel-mode Lorentz microscopy of one single dot of thickness
  $t=\lengthnm{60\pm20}$ as a function of rising (i.-vi.) and then decreasing (vi.-viii.) field.
  (field of view $\lengthnm{1100\times300}$)). The in-plane
  component of the applied field is indicated on each image. The location of the
  surface vortex associated with the switching of a N\'{e}el cap is indicated with an arrow. (b)~Simulation of Fresnel
  mode images of the switch of N\'{e}el cap, based on the micromagnetic configurations of
  \subfigref{fig-hysteresis}b i., iii. and iv. The defocusing is $\lengthmicron{500}$}
  \end{center}
\end{figure}

The dots were then investigated under an \insitu magnetic field by Lorentz microscopy. In the Fresnel mode the domain wall is revealed by fringes arising from interferences of
electrons having gone through the two longitudinal domains\bracketsubfigref{fig-lorentz}a.
Computed Fresnel contrasts based on simulated micromagnetic configurations show that domain walls
with parallel NCs give rise to a narrower pattern of fringes\bracketsubfigref{fig-lorentz}b,
allowing us to follow in the experiments the location of the propagating surface vortex
responsible for the switching of a NC. All features predicted by
the simulations\bracketsubfigref{fig-hysteresis}b are thus confirmed: the hysteresis~(see \eg iv.
versus viii.); the translational susceptibility of the surface vortex that is negligible for low
field and increases dramatically close to the vicinity of the switching field upon increasing the
field~(i.-v., compare with \subfigref{fig-hysteresis}{b i.-iii.}); the negligible or even zero
susceptibility before the switching back when the field is decreased~(vi.-vii., compare with
\subfigref{fig-hysteresis}{b iv.-v.}). Besides as opposite chiralities induce dark or light
fringes respectively, the experimental images (i.-viii.) taken through the hysteresis demonstrate
that the chirality is not affected by the switching of the NCs, neither at rising nor at
decreasing field. This is consistent with XMCD-PEEM experiments where the occurrence of clockwise
and anticlockwise chiralities remained similar for all applied magnetic fields and no
cross-correlation between the chirality and the state of the NCs could be evidenced either, within
the statistical error bars\cite{bib-FRU2008}. Concerning the polarity of the DW core, which is
detectable neither by XMCD-PEEM nor by Lorentz microscopy, we must rely only on the simulations
to infer that the core of the DW remains unaffected by the switch of NCs.

To conclude we have demonstrated numerically and experimentally that the direction of
magnetization of the N\'{e}el caps, an internal surface feature of Bloch walls, can be switched
controllably using a transverse magnetic field. This was achieved in elongated ferromagnetic dots
displaying a flux-closure state, so that the domain wall can be manipulated while the domains
remain unaffected owing to the internal demagnetizing field\cite{bib-NEELCAPS-arrott}. The switching occurs through the
propagation of surface vortices through the NCs. The selection of the state of the NCs at
remanence is made possible by the tilted facets of the dots, which lift the degeneracy between the
top and bottom surfaces. The switching affects neither the chirality of the in-plane domains nor
the polarity of the core of the DW. These results show that at least three degrees of freedom may be
addressed independently and reliably in a single magnetic dot, whereas only two had been manipulated so far in
the now widely-studied vortex state found in high-symmetry dots. This fundamental knowledge of the
possibility of an internal switching in Bloch walls may also be of use for studies of field- or
current-driven movement of domain walls, were it is well known that domain walls can undergo
topological transformations during their movement\cite{bib-KLA2006}.

\section*{References}


\end{document}